\documentclass[a4paper,12pt]{article}
\usepackage{amssymb}
\usepackage{graphicx}
\usepackage{epsfig}
\usepackage{booktabs}
\usepackage{array}
\usepackage{multirow}
\usepackage{hyperref}
\usepackage{cite}
\usepackage{appendix}
\usepackage{amsmath}
\begin{document}
\title{Self-similarity and the Froissart bound}
\author{ $ \mathrm{Akbari \; Jahan}^{\star} $ and D K Choudhury\\ Department of Physics, Gauhati University,\\ Guwahati - 781 014, Assam, India.\\ $ {}^{\star} $Email: akbari.jahan@gmail.com}
\date{}
\maketitle
\begin{abstract}
The Froissart bound implies that the total cross section (or, equivalently, the structure function) cannot rise faster than the logarithmic growth of $ \ln^{2} \left( \frac{1}{x} \right)$. In this work, we show that such a slow growth is not compatible with the notion of self-similarity. As a result, it calls for the modification of the defining transverse-momentum-dependent parton density function (TMD PDF) of a self-similarity based proton structure function $F_{2} \left(x,Q^{2} \right)$ at small \textit{x}. Using plausible assumptions, we obtain the Froissart saturation condition on this TMD PDF.\\

\textbf{Keywords \,:} Froissart bound, self-similarity, TMD PDF.\\

\end{abstract}
\section{Introduction}
The notion of fractals found its applicability in high energy physics through the self-similar nature of hadron multiparticle production processes \cite{1,2,3,4}. The fractal characters of hadrons had also been pursued within a specific quark model \cite{5}. Dremin and Levtchenko \cite{6} first noted the relevance of these ideas in the contemporary physics of deep inelastic scattering. However, it was Lastovicka \cite{7} who brought further attention to this notion when he proposed a relevant formalism and suggested a functional form of the structure function $F_{2}\left(x,Q^{2}\right)$ valid at small \textit{x}. In recent years, the formalism was further analyzed phenomenologically \cite{8,9}.\\

In the present work, we explore the possibility of Froissart saturation \cite{10} in the self-similarity based model of the proton, as it has attracted attention in the recent literature \cite{11,12,13,14,15,16}. It is well known that in the conventional QCD evolution equations, like the Dokshitzer-Gribov-Lipatov-Altarelli-Parisi (DGLAP) \cite{17,18,19} and Balitsky-Fadin-Kuraev-Lipatov (BFKL) \cite{20,21,22,23} approaches, this limit is violated; while in the DGLAP approach the small-\textit{x} gluons grow faster than any power of $\displaystyle \ln \frac{1}{x} \approx \ln \left(\frac{s}{Q^{2}} \right) $ \cite{24}, in the BFKL approach it grows as a power of $ \displaystyle \frac{1}{x}$ \cite{20,21,22,23,24,25}.\\

In Sec. 2 we outline the formalism, and Sec. 3 gives the results and a discussion. The conclusions of this work are highlighted in Sec. 4.

\section{Formalism}
\subsection{TMD PDF in the self-similarity based model}
The transverse-momentum-dependent parton density function (TMD PDF), having the notion of self-similarity at small \textit{x}, is defined as \cite{7}
\begin{eqnarray}
\label{eqn:logfi}
\ln f_{i}\left( x,k_{t}^{2}\right)& = & D_{1}\ln \left( \frac{1}{x} \right)\ln \left(1+\frac{k_{t}^{2}}{Q_{0}^{2}}\right)+D_{2}\ln \left(\frac{1}{x}\right)+ \nonumber \\
& & D_{3}\ln \left( 1+\frac{k_{t}^{2}}{Q_{0}^{2}}\right)+D_{0}^{i}-\ln M^{2} \, ,
\end{eqnarray}
so that
\begin{equation}
\label{eqn:fi}
f_{i}\left(x,k_{t}^{2}\right) = \left(1+ \frac{k_{t}^{2}}{Q_{0}^{2}} \right)^{D_{3}+D_{1}\ln \left(\frac{1}{x}\right)} \left( \frac{1}{x}\right)^{D_{2}} \frac{e^{D_{0}^{i}}}{M^{2}} \, ,
\end{equation}
and the quark density for the \textit{i} th flavor is
\begin{equation}
\label{eqn:qi_int}
q_{i}\left(x,Q^{2}\right) = \int\limits_{0}^{Q^{2}}\, dk_{t}^{2} \, f_{i}\left(x,k_{t}^{2}\right) \, ,
\end{equation}
i.e.,
\begin{equation}
\label{eqn:qi}
q_{i}\left(x,Q^{2}\right) = \frac{e^{D_{0}^{i}}}{M^{2}} \, \frac{Q_{0}^{2} \left( \frac{1}{x}\right)^{D_{2}}}{1+D_{3}+D_{1} \ln \left( \frac{1}{x}\right)} \left( \left( \frac{1}{x}\right)^{D_{1} \ln \left( 1+ \frac{Q^{2}}{Q_{0}^{2}}\right)} \left( 1+\frac{Q^{2}}{Q_{0}^{2}}\right)^{D_{3}+1} -1\right) \, ,
\end{equation}
and the structure function $ F_{2} \left( x,Q^{2} \right) $ is defined as
\begin{equation}
\label{eqn:SF_expression}
F_{2}\left(x,Q^{2} \right) = x \, \sum_{i} e_{i}^{2} \left(q_{i} \left(x,Q^{2}\right)+ \bar{q_{i}}\left(x,Q^{2}\right) \right).
\end{equation}
The proton structure function thus obtained is
\begin{equation}
\label{eqn:F2}
F_{2}\left(x,Q^{2} \right) =  \frac{e^{D_{0}}}{M^{2}} \left\lbrace \frac{Q_{0}^{2} \, \left( \frac{1}{x}\right)^{D_{2}} \, x }{1+D_{3}+D_{1} \ln \left(\frac{1}{x} \right)} \left( \left( \frac{1}{x} \right)^{D_{1} \ln \left( 1+ \frac{Q^{2}}{Q_{0}^{2}}\right)} \left( 1+\frac{Q^{2}}{Q_{0}^{2}}\right)^{D_{3}+1}-1 \right) \right\rbrace \, , 
\end{equation}
where $ D_{0}=\sum \limits_{i} D_{0}^{i} $, and
\begin{eqnarray}
\label{eqn:parameters}
D_{0} & = & 0.339 \pm 0.145 \, , \nonumber \\
D_{1} & = & 0.073\pm 0.001 \, , \nonumber \\
D_{2} & = & 1.013\pm 0.01 \, , \nonumber \\
D_{3} & = & -1.287\pm 0.01 \, , \nonumber \\
Q_{0}^{2} & = & 0.062\pm 0.01 \; \mathrm{GeV}^{2}.
\end{eqnarray}
The above parameters have been fitted from HERA data \cite{26,27} in the range $ 6.2\, \times \, 10^{-7} \leq x \leq 10^{-2} $ and $ 0.045 \leq Q^{2} \leq 120 \, \mathrm{GeV^{2}}$.\\

An additional term, $ -\ln M^{2}$ with $ M^{2} $, having dimension of energy squared is introduced in Eq. (\ref{eqn:logfi}) so that the integrated parton distribution function defined in Eq. (\ref{eqn:qi_int}) is dimensionless. We set $ M^{2}\simeq 1 \, \mathrm{GeV}^{2} $.

\subsection{Incorporation of the Froissart bound in TMD PDF}
The Froissart bound \cite{10} is a well-established property of the strong interactions and puts a strict limit on the rate of growth with energy of the total cross sections. It implies that the total cross section (or, equivalently, the structure function) cannot rise faster than the logarithmic growth of $ \ln^{2}\left(\frac{1}{x}\right)$. Such a slow logarithmic growth is however not compatible with the model of Ref.\cite{7}, which instead has a power law growth in $1/x$.\\

In order to incorporate the Froissart bound into the model, we introduce the hard scale $ Q^{2} $ and an additional function $ h\left(x, k_{t}^{2}, Q^{2} \right)$ in the defining TMD PDF [Eq. (\ref{eqn:logfi})],
\begin{eqnarray}
\label{eqn:logfi_ext}
\ln f_{i} \left(x, k_{t}^{2},Q^{2} \right) & = & D_{1}\ln \left(\frac{1}{x} \right)\ln \left( 1+ \frac{k_{t}^{2}}{Q_{0}^{2}}\right)+ D_{2} \ln \left(\frac{1}{x} \right)+  \nonumber \\
& & D_{3} \ln \left( 1+ \frac{k_{t}^{2}}{Q_{0}^{2}}\right)+ D_{0}^{i}- \ln M^{2}+ \ln h \left(x, k_{t}^{2}, Q^{2} \right) \, , \nonumber \\
\end{eqnarray}
so that
\begin{equation}
\label{eqn:fi_ext}
f_{i}\left(x,k_{t}^{2},Q^{2} \right)= \left( 1+ \frac{k_{t}^{2}}{Q_{0}^{2}}\right)^{D_{3}+D_{1} \ln \left(\frac{1}{x} \right)}\left( \frac{1}{x}\right)^{D_{2}} \frac{e^{D_{0}^{i}}}{M^{2}}\, h \left( x,k_{t}^{2},Q^{2}\right).
\end{equation}
The integrated quark density then becomes
\begin{equation}
\label{eqn:qi_ext}
q_{i}\left( x,Q^{2}\right)= \int \limits_{0}^{Q^{2}} \, dk_{t}^{2}\left( 1+ \frac{k_{t}^{2}}{Q_{0}^{2}}\right)^{D_{3}+D_{1} \ln \left( \frac{1}{x}\right)} \left( \frac{1}{x}\right)^{D_{2}} \frac{e^{D_{0}^{i}}}{M^{2}} \, h \left( x,k_{t}^{2}, Q^{2} \right).
\end{equation}
Since the explicit $ k_{t}^{2} $ dependence of the function $ h \left( x,k_{t}^{2}, Q^{2} \right) $ is necessary to evaluate the integral over $ k_{t}^{2} $, we assume that $ h \left( x,k_{t}^{2}, Q^{2} \right) $ is factorizable in $ k_{t}^{2} $ and \textit{x} as
\begin{equation}
\label{eqn:hxkt2Q2}
h \left( x,k_{t}^{2},Q^{2}\right)= h \left(x,Q^{2} \right) h \left( k_{t}^{2}\right).
\end{equation}
Specifically, we assume that $ h \left( k_{t}^{2}\right) $ is normalized as
\begin{equation}
\label{eqn:h_norm}
\int h \left( k_{t}^{2}\right) \, dk_{t}^{2} = 1 \, ,
\end{equation}
and that it has the form suggested by Zavada \cite{28},
\begin{equation}
\label{eqn:hkt2}
h \left( k_{t}^{2}\right)= \frac{1}{\langle k_{t}^{2} \rangle } e^{\left( \frac{-k_{t}^{2}}{\langle k_{t}^{2} \rangle} \right) } \, ,
\end{equation}
with the average of $ k_{t}^{2} $, $ \langle k_{t}^{2} \rangle = 0.25 \, \mathrm{GeV^{2}}$ \cite{29}.\\

Using Eqs. (\ref{eqn:hxkt2Q2}) and (\ref{eqn:hkt2}) in Eq. (\ref{eqn:qi_ext}), we have
\begin{equation}
\label{eqn:qi_final}
q_{i}\left( x,Q^{2}\right)= \frac{e^{D_{0}^{i}}}{M^{2}} \, \left(\frac{1}{x} \right)^{D_{2}} \, \frac{1}{\langle k_{t}^{2} \rangle} \, h \left(x,Q^{2} \right) \, \int \limits_{0}^{Q^{2}} \, dk_{t}^{2}\left( 1+ \frac{k_{t}^{2}}{Q_{0}^{2}}\right)^{D_{3}+D_{1} \ln \left( \frac{1}{x} \right)} \, e^{\left( \frac{-k_{t}^{2}}{\langle k_{t}^{2} \rangle} \right) }.
\end{equation}

Using Eq. (\ref{eqn:qi_final}) in Eq. (\ref{eqn:SF_expression}), we get the modified version of the structure function as
\begin{equation}
\label{eqn:F2_expression}
\hat{F_{2}} \left( x,Q^{2}\right)= \frac{e^{D_{0}}}{M^{2}} \, \left(\frac{1}{x} \right)^{D_{2}-1} \, \frac{1}{\langle k_{t}^{2} \rangle} \, h \left(x,Q^{2} \right) \, I \left(x,Q^{2} \right) \, ,
\end{equation}
where
\begin{equation}
\label{eqn:I}
I \left(x,Q^{2} \right)=\int \limits_{0}^{Q^{2}} \, dk_{t}^{2}\left( 1+ \frac{k_{t}^{2}}{Q_{0}^{2}}\right)^{D_{3}+D_{1} \ln \left( \frac{1}{x} \right)} \, e^{\left( \frac{-k_{t}^{2}}{\langle k_{t}^{2} \rangle}\right)}.
\end{equation}
As mentioned earlier, the Froissart saturation condition is obtained if
\begin{equation}
\label{eqn:F2_condiiton}
\hat{F_{2}}\left( x,Q^{2}\right)\leq \ln ^{2} \frac{1}{x} \, .
\end{equation}
Equating the rhs of Eq. (\ref{eqn:F2_expression}) and Eq. (\ref{eqn:F2_condiiton}), we have
\begin{equation}
\label{eqn:h_condition}
h \left( x,Q^{2} \right)\leq \frac{M^{2} e^{-D_{0}} \ln^{2} \left( \frac{1}{x}\right) \langle k_{t}^{2}\rangle }{\left( \frac{1}{x} \right)^{D_{2}-1} \, I \left( x,Q^{2} \right)}.
\end{equation}
Equation (\ref{eqn:h_condition}) is the desired Froissart saturation condition on the TMD PDF in the present approach. It implies that a Froissart-compatible $h\left(x,Q^{2} \right)$ should be able to cancel the power law divergent $\left( \frac{1}{x}\right)^{D_{2}-1}$ factor as well as cancel the effect of the generalized incomplete gamma function $I \left(x,Q^{2} \right)$, where the \textit{x} dependence comes from the exponent $D_{3}+D_{1} \ln \left(\frac{1}{x} \right)$.\\

Note that the function $h \left(x,k_{t}^{2},Q^{2} \right)$ is dimensionless and $h \left(x,k_{t}^{2},Q^{2} \right)=1$ in the model \cite{7}.

\section{Results and discussion}
For the Froissart saturation to be incorporated into the modified model of Eq. (\ref{eqn:F2_expression}), we need to choose a model fit to the data that has the required $ \ln^{2} \left( \frac{1}{x}\right) $ behavior and then appropriately modify the \textit{x} dependence of the model of Ref.\cite{7}. To this end, we use the expression for the proton structure function of Ref. \cite{14},
\begin{eqnarray}
\label{eqn:F2_Block}
F_{2}^{p} \left(x,Q^{2} \right)=\left(1-x \right)\left\lbrace \frac{F_{p}}{1-x_{p}}+A(Q^{2})\ln \frac{x_{p} \left(1-x \right)}{x \left( 1-x_{p}\right)}+B(Q^{2})\ln^{2} \frac{x_{p} \left(1-x \right)}{x \left( 1-x_{p}\right)} \right\rbrace \, , \nonumber \\
\end{eqnarray}
where
\begin{eqnarray}
A(Q^{2})=a_{0}+a_{1}\ln Q^{2}+a_{2} \ln^{2} Q^{2} \, , \nonumber \\
B(Q^{2})=b_{0}+b_{1}\ln Q^{2}+b_{2} \ln^{2} Q^{2} \, ,
\end{eqnarray}
and the parameters fitted from deep inelastic scattering data \cite{14} are
\begin{equation}
x \leq x_{p}=0.11 \quad \mathrm{and} \quad F_{p}=0.413 \pm 0.003 \, , \nonumber \\
\end{equation}
\begin{eqnarray}
a_{0} & = & -8.471 \times 10^{-2} \pm 2.62 \times 10^{-3} \, , \nonumber \\
a_{1} & = & 4.190 \times 10^{-2} \pm 1.56 \times 10^{-3} \, , \nonumber \\
a_{2} & = & -3.976 \times 10^{-3} \pm 2.13 \times 10^{-4} \, , \nonumber \\
b_{0} & = & 1.292 \times 10^{-2} \pm 3.62 \times 10^{-4} \, , \nonumber \\
b_{1} & = & 2.473 \times 10^{-4} \pm 2.46 \times 10^{-4} \, , \nonumber \\
b_{2} & = & 1.642 \times 10^{-3} \pm 5.52 \times 10^{-5}\, .
\end{eqnarray}
The explicit form of the function $h\left(x,Q^{2}\right)$ using Ref.\cite{14} is given as
\begin{equation}
\label{eqn:h_Froissart}
h\left(x,Q^{2}\right)\leq \frac{M^{2}}{e^{D_{0}}}\left(\frac{1}{x}\right)^{1-D_{2}}\frac{\langle k_{t}^{2} \rangle}{I\left(x,Q^{2}\right)} \, F_{2}^{p}\left(x,Q^{2} \right) \, ,
\end{equation}
where $I\left(x,Q^{2}\right)$ is as given in Eq. (\ref{eqn:I}) and $F_{2}^{p}\left(x,Q^{2} \right)$ is computed using the expression for the proton structure function as given in Eq. (\ref{eqn:F2_Block}). We note that, by construction, the model of Ref. \cite{14} has Froissart saturation for $ x\leqslant 0.09 $, which is well within the valid kinematic region of the present model \cite{7}.\\

Equation (\ref{eqn:h_Froissart}), when used in Eq. (\ref{eqn:F2_expression}), will have the desired Froissart saturation in the modified version of the model. $h\left(x,Q^{2}\right)$ of Eq. (\ref{eqn:F2_expression}) will then appropriately modify the \textit{x} dependence of the model of Ref. \cite{7}. Evaluating the integral $ I \left(x,Q^{2}\right) $ of Eq. (\ref{eqn:I}) numerically and taking $ \langle k_{t}^{2} \rangle = 0.25 \, \mathrm{GeV^{2}}$ \cite{29}, we compute the values of $h \left(x,Q^{2} \right)$ in the region $ 6.2\, \times \, 10^{-7} \leq x \leq 10^{-2} $ and $ 0.045 \leq Q^{2} \leq 120 \, \mathrm{GeV^{2}}$ and tabulate them in Table \ref{tab:1}.

\begin{table}[!h]
\begin{center}
\caption{\textbf{Values of $h\left(x,Q^{2}\right)$ for given sets of \textit{x} and $Q^{2}$ using the model of Ref.\cite{14}.}}
\label{tab:1}
\bigskip
\begin{tabular}{|c|c|c|c|c|c|c|}
\hline
$Q^{2}\left( \mathrm{GeV^{2}} \right)$ & \multicolumn{6}{c}{$h \left(x,Q^{2} \right) \, \left(\mathrm{in \, GeV^{2}} \right)$} \vline  \\ \cline{2-3} \cline{4-5} \cline{6-7}
{} & $x=10^{-2}$   &  $x=10^{-3}$ & $x=10^{-4}$ & $x=10^{-5}$ & $x=10^{-6}$ & $x=6.2 \times 10^{-7}$ \\ \hline
10 & 1.1051 & 1.4917 & 1.9973 & 2.4930 & 2.8903 & 2.9554 \\ 
30 & 1.3321 & 2.0453 & 2.8966 & 3.7004 & 4.3325 & 4.4352 \\ 
45 & 1.4162 & 2.2681 & 3.2724 & 4.2162 & 4.9574 & 5.0779 \\ 
60 & 1.4761 & 2.4322 & 3.5535 & 4.6051 & 5.4311 & 5.5656 \\ 
120 & 1.6207 & 2.8481 & 4.2796 & 5.6203 & 6.6758 & 6.8483 \\ \hline
\end{tabular}
\end{center}
\end{table}
Table \ref{tab:1} shows that the modification is modest within the realm of the \textit{x} and $Q^{2}$ values studied, which is reasonable, since similar data is successfully fit by both models. Large corrections are needed only at extremely small \textit{x} and large $Q^{2}$.\\

\begin{figure}[h]
\begin{minipage}{18pc}
\includegraphics[scale=.45]{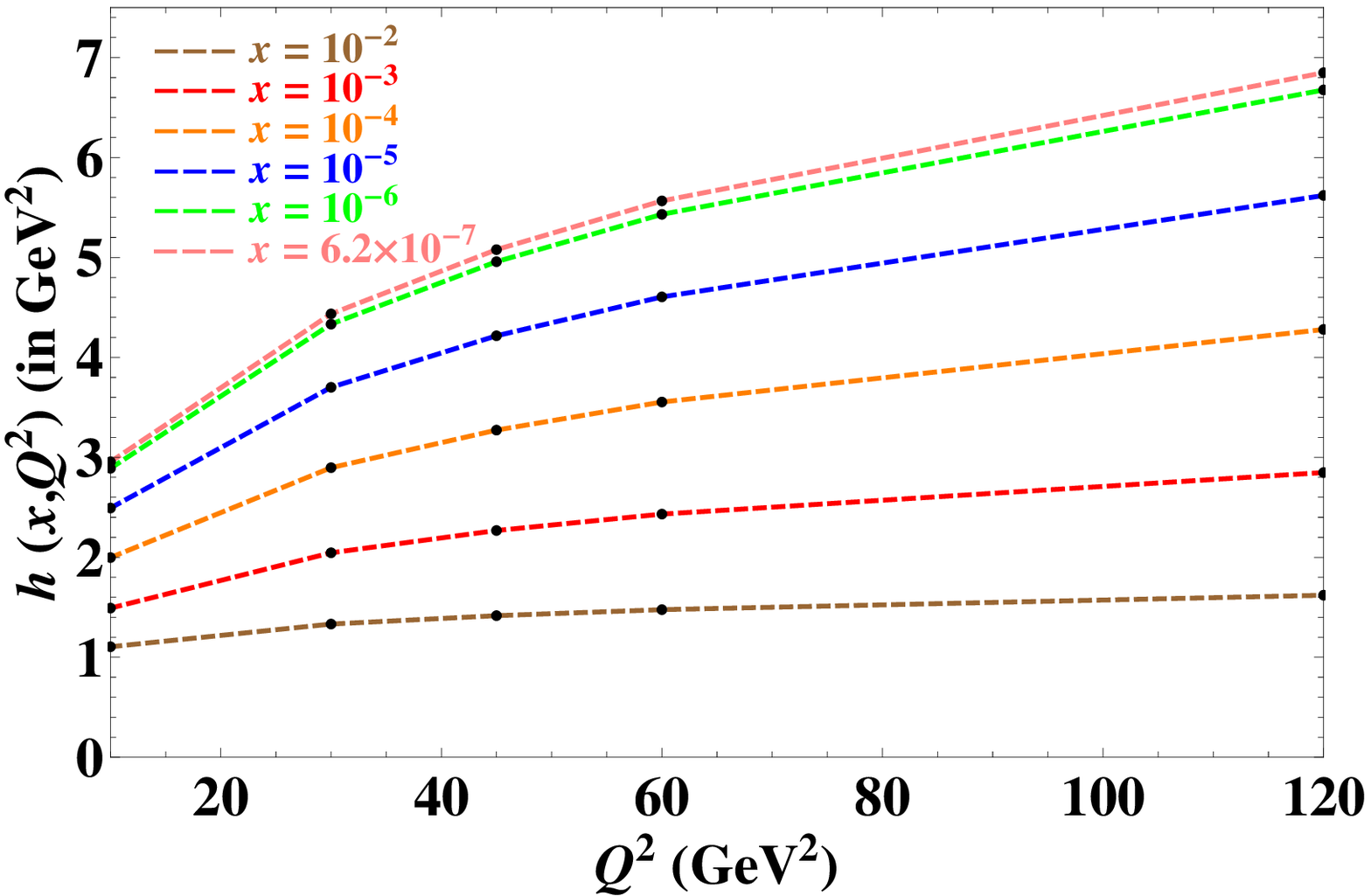}
\caption[$h \left(x,Q^{2} \right)$ vs $Q^{2}$ using the model of Block \textit{et al}.]{$h \left(x,Q^{2} \right)$ vs $Q^{2}$ using the model of Ref.\cite{14}.}
\label{fig:1}
\end{minipage}
\hspace{1.5pc}
\begin{minipage}{18pc}
\includegraphics[scale=.45]{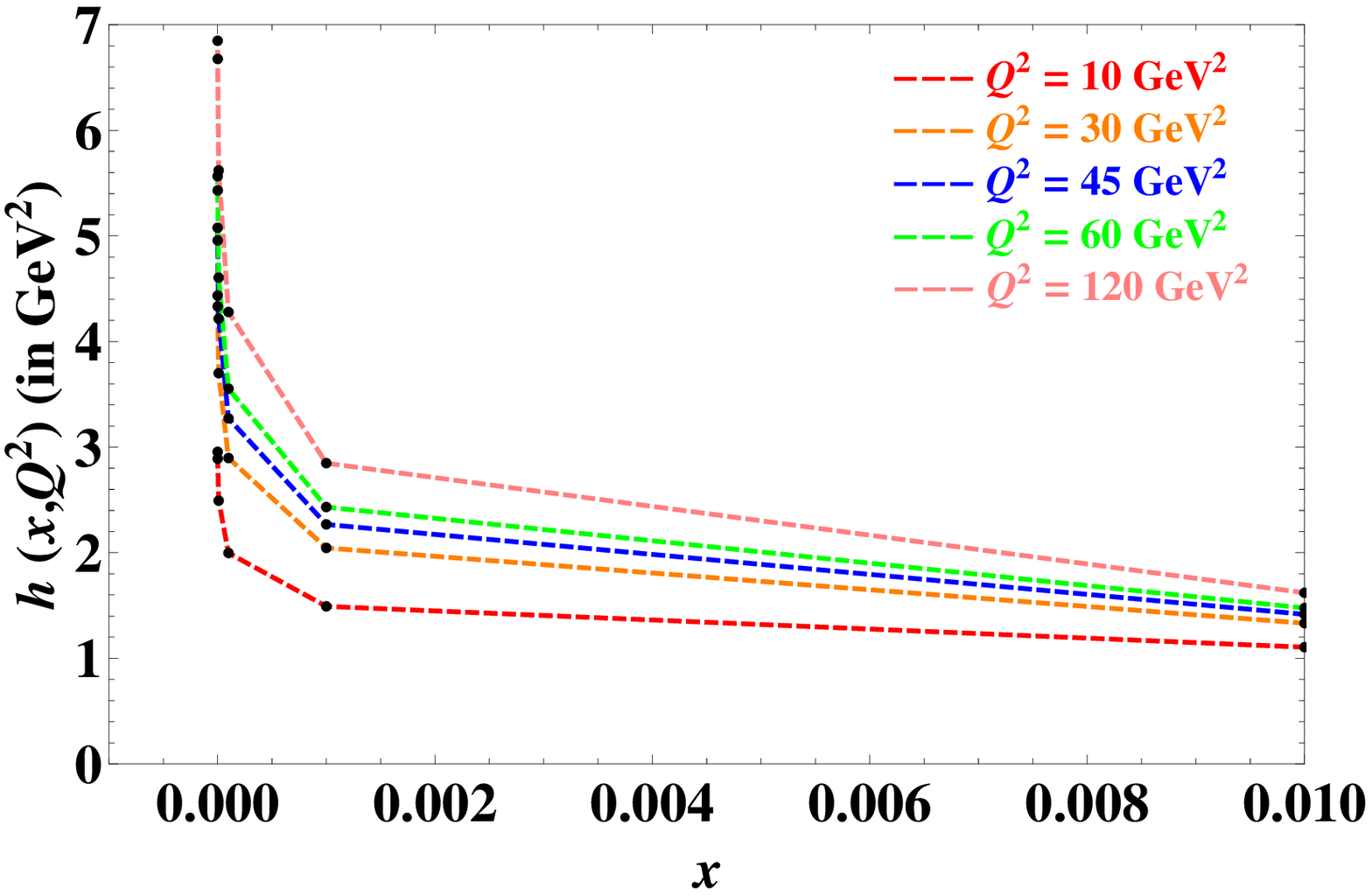}
\caption[$h \left(x,Q^{2} \right)$ vs $x$ using the model of Block \textit{et al}.]{$h \left(x,Q^{2} \right)$ vs $x$ using the model of Ref.\cite{14}.}
\label{fig:2}
\end{minipage}
\end{figure}

Graphical representations of the function $h \left(x,Q^{2} \right)$ vs $Q^{2}$ and $h \left(x,Q^{2} \right)$ vs \textit{x} are shown in Figure \ref{fig:1} and Figure \ref{fig:2} respectively.\\

The figures show that $h \left(x,Q^{2} \right)$ increases with $Q^{2}$ (Figure \ref{fig:1}), while it decreases with \textit{x} (Figure \ref{fig:2}). But the growth with $Q^{2}$ is much slower than that of the model of Ref.\cite{7}, as expected.

\section{Conclusions}
In the present work, we have incorporated the transverse momentum dependence in the self-similarity based model of the proton at small $ \textit{x}$. We then argued that the logarithmic growth in $\ln^2 \left( \frac{1}{x} \right)$ is not compatible with the original self-similarity based model of Ref.\cite{7}. We therefore added an additional function $ h \left( x,k_{t}^{2},Q^{2}\right)$ containing two hard scales in the defining TMD PDF and obtained its compatibility with the Froissart bound. Comparing it with the phenomenological Froissart saturation model of Ref.\cite{14}, we obtained the modified \textit{x} dependence of the structure function through Eq. (\ref{eqn:h_Froissart}).\\

It is also to be noted that the case $D_{2} \approx 1$, considered in Ref. \cite{7}, fits just as well as the case when $D_{2}$ is allowed to vary in the fit to the data, so the essential \textit{x} dependence in $F_{2}\left(x,Q^{2} \right)$ comes from the $D_{1}$ term in the model corresponding to the correlation term of the TMD of Eq. (\ref{eqn:logfi}).\\

Plausible dynamics beneath the Froissart bound have already been suggested in the literature \cite{30,31}. It is also well known that if a new process like gluon recombination starts \cite{32}, the number of small-\textit{x} gluons might saturate to a limit compatible with the Froissart bound. The model of Ref. \cite{7} falls short of accommodating such dynamics explicitly, as does the present work. However, the introduction of the new function $ h \left( x,Q^{2}\right)$ [with the explicit representation given by Eq. (\ref{eqn:h_Froissart})] indicates a plausible way of parametrizing such effects in the modified version of the model. The physical significance of $ h \left( x,Q^{2}\right)$ is just this.\\

In this work, we have taken the notion of self-similarity merely as a basis to parametrize the TMD PDF, the parton distribution function, and eventually the structure function $F_{2}\left(x,Q^{2} \right)$ at small \textit{x}. However, the variable in which the supposed fractal scaling of the quark distribution and $F_{2} \left( x,Q^{2} \right)$ occurs is not known from the underlying theory. In Ref.\cite{7}, the choice of $1/x$ is presumably because of the power-law form of the quark distribution at small \textit{x} found in the Gl\"{u}ck-Reya-Vogt (GRV) distribution \cite{33}. However, this form is not derived theoretically or from data, but rather follows from the power-law distribution in \textit{x} assumed for the input quark distribution used by the GRV distribution \cite{33} for the QCD evolution. The choice of $1/x$ as the proper scaling variable is therefore not established from the underlying theory.\\

A more plausible variable appears instead to be $\ln \, (1/x)$ (as has been used in the function of Ref.\cite{14}), which is consistent with the behavior of very high-energy interactions. In this case, when the self-similar TMD $f_{i} \left(x,k_{t}^{2} \right)$ of quark flavor \textit{i} is defined by the simple scaling variable $\ln \, (1/x)$, it will have the form
\begin{equation}
\label{eqn:fi_prime}
\ln f_{i}^{\prime}\left(x\right)=D_{i} \, \ln \left(\ln \frac{1}{x} \right) \, ,
\end{equation}
causing the PDF defined in Eq. (\ref{eqn:qi_int}) to be
\begin{equation}
\label{eqn:qi_prime}
q_{i}^{\prime} \left(x,Q^{2} \right)= \left[\ln \left( \frac{1}{x}\right) \right]^{D_{i}} \, Q^{2} \, ,
\end{equation}
where $D_{i}$ is the proportionality constant.\\

The PDF and hence the structure function will be trivially compatible with the Froissart bound $\ln^{2} \left( \frac{1}{x}\right)$ with $D_{i} = 2$.\\

A similar replacement in Lastovicka's parametrization with two magnification factors $\ln \left( \frac{1}{x} \right)$ and $\left(1+\frac{k_{t}^{2}}{Q_{0}^{2}} \right)$ would result in
\begin{equation}
\label{eqn:fi_tilde}
\tilde{f}_{i}\left(x,k_{t}^{2}\right) = \left(1+ \frac{k_{t}^{2}}{Q_{0}^{2}} \right)^{D_{3}+D_{1}\ln \left(\ln \frac{1}{x}\right)} \left(\ln \frac{1}{x}\right)^{D_{2}} \frac{e^{D_{0}^{i}}}{M^{2}} \, ,
\end{equation}
and the subsequent PDF becomes
\begin{equation}
\label{eqn:qi_tilde}
\tilde{q}_{i}\left(x,Q^{2}\right) = \frac{e^{D_{0}^{i}} \, Q_{0}^{2}}{M^{2}} \, \frac{\left(\ln \frac{1}{x}\right)^{D_{2}}}{1+D_{3}+D_{1} \ln \left(\ln \frac{1}{x}\right)} \left( \left(\ln \frac{1}{x}\right)^{D_{1} \ln \left( 1+ \frac{Q^{2}}{Q_{0}^{2}}\right)} \left( 1+\frac{Q^{2}}{Q_{0}^{2}}\right)^{D_{3}+1} -1\right).
\end{equation}
For very small \textit{x} and large $Q^{2}$, the second term of Eq. (\ref{eqn:qi_tilde}) can be neglected, leading to
\begin{equation}
\label{eqn:qi_tilde_smallx}
\tilde{q}_{i}\left(x,Q^{2}\right) = \frac{e^{D_{0}^{i}} \, Q_{0}^{2}}{M^{2}} \, \frac{\left(\ln \frac{1}{x}\right)^{D_{2}+D_{1} \, \ln \left(1+ \frac{Q^{2}}{Q_{0}^{2}} \right)}}{1+D_{3}+D_{1} \ln \left(\ln \frac{1}{x}\right)} \left( 1+\frac{Q^{2}}{Q_{0}^{2}}\right)^{D_{3}+1} \, ,
\end{equation}
which satisfies the Froissart saturation condition if
\begin{equation}
\label{eqn:28}
D_{2}+D_{1} \, \ln \left(1+ \frac{Q^{2}}{Q_{0}^{2}} \right) = 2
\end{equation}
within $\ln\ln \left( \frac{1}{x} \right)$ corrections. The structure function will also have a similar result.\\

Thus the original model of Ref.\cite{7} has two features:
\begin{enumerate}
\item The Froissart saturation condition is satisfied only with $\ln\ln \left( \frac{1}{x} \right)$ corrections.
\item The parameters $D_{1}$ and/or $D_{2}$ themselves should be $Q^{2}$ dependent so that the rhs of Eq. (\ref{eqn:28}) is $Q^{2}$ independent.
\end{enumerate}
In order to obtain the exact Froissart saturation condition without $\ln\ln \left(\frac{1}{x} \right)$ corrections, the correlation parameter $D_{1}$ should be vanishingly small $(D_{1} \approx 0)$, in which case the lhs of Eq. (\ref{eqn:28}) will be $Q^{2}$ independent as well.\\

On the other hand, it could also be argued that in order to have Froissart-like behavior the confinement scale should enter in the description. However, the notion of confinement is completely absent in the present approach. One possible way to incorporate such a scale is presumably through the redefinition of the scaling variable involving virtuality $Q^{2}$, as has been done in Ref.\cite{34} for the running coupling constant $\alpha_{s}(Q^{2})$. It will be interesting to study the consequences of the model in the future by suitably incorporating the confinement scale as well.


\begin{thebibliography}{70}
\bibitem{1} I. M. Dremin, \textit{Mod. Phys. Lett.} \textbf{A3}, 1333 (1988).
\bibitem{2} R. Lipa and B. Buschbeck, \textit{Phys. Lett.} \textbf{B223}, 465 (1989).
\bibitem{3} W. Florkowski and R. C. Hwa, \textit{Phys. Rev.} \textbf{D43}, 1548 (1991).
\bibitem{4} D. Ghosh, A. Deb, S. R. Sahoo, P. K. Haldar and M. Mondal, \textit{Europhys. Lett.} \textbf{65}, 472 (2004)
\bibitem{5} S. N. Banerjee, A. Bhattacharya, B. Chakrabarti and S. Banerjee, \textit{Int. J. Mod. Phys.} \textbf{A16}, 201 (2001); \textit{ibid.} \textbf{A17}, 4939 (2002).
\bibitem{6} I. M. Dremin and B. B. Levtchenko, \textit{Phys. Lett.} \textbf{B292}, 155 (1992).
\bibitem{7} T. Lastovicka, \textit{Euro. Phys. J.} \textbf{C24}, 529 (2002); \textbf{arXiv:}hep-ph/0203260.
\bibitem{8} D. K. Choudhury and R. Gogoi; \textit{Ind. J. Phys.} \textbf{80}, 659 (2006); \textbf{arXiv:}hep-ph/0503047.
\bibitem{9} Akbari Jahan and D. K. Choudhury; \textit{Ind. J. Phys.} \textbf{85}, 587 (2011); \textbf{arXiv:}1102.0069[hep-ph].
\bibitem{10} M. Froissart, \textit{Phys. Rev.} \textbf{123}, 1053 (1961).
\bibitem{11} M. M. Block, E. L. Berger and C. I. Tan, \textit{Phys. Rev. Lett.} \textbf{97}, 252003 (2006); \textbf{arXiv:}hep-ph/0610296.
\bibitem{12} E. L. Berger, M. M. Block and C. I. Tan, \textit{Phys. Rev. Lett.} \textbf{98}, 242001 (2007); \textbf{arXiv:}hep-ph/0703003.
\bibitem{13} F. Carvalho, F. O. Duraes, V. P. Goncalves and F. S. Navarra, \textit{Mod. Phys. Lett.} \textbf{A23}, 2847 (2008); \textbf{arXiv:}0705.1842 [hep-ph].
\bibitem{14} M. M. Block, L. Durand, P. Ha and D. W. McKay, \textit{Phys. Rev.} \textbf{D84}, 094010 (2011); \textbf{arXiv:}1108.1232[hep-ph].
\bibitem{15} Y. Azimov, \textbf{arXiv:}1208.4304[hep-ph].
\bibitem{16} A. Martin and S. M. Roy, \textit{preprint} CERN-PH-TH/2013-113 (2013); \textbf{arXiv:}1306.5210[hep-ph].
\bibitem{17} G. Altarelli and G. Parisi, \textit{Nucl. Phys.} \textbf{B126}, 298 (1977).
\bibitem{18} Y. L. Dokshitzer, \textit{Sov. Phys. JETP} \textbf{46}, 641 (1977).
\bibitem{19} V. N. Gribov and L. N. Lipatov, \textit{Sov. J. Nucl. Phys.} \textbf{15}, 438 (1972).
\bibitem{20} V. S. Fadin, E. A. Kuraev and L. N. Lipatov, \textit{Phys. Lett.} \textbf{B60}, 50 (1975).
\bibitem{21} E. A. Kuraev, L. N. Lipatov and V. S. Fadin, \textit{Sov. Phys. JETP} \textbf{44}, 443 (1976).
\bibitem{22} E. A. Kuraev, L. N. Lipatov and V. S. Fadin, \textit{Sov. Phys. JETP} \textbf{45}, 199 (1977).
\bibitem{23} I. I. Balitsky and L. N. Lipatov, \textit{Sov. J. Nucl. Phys.} \textbf{28}, 822 (1978).
\bibitem{24} R. G. Roberts, \textit{The Structure of Proton}, Cambridge University Press, Cambridge, p30 (1990).
\bibitem{25} L. N. Lipatov, \textit{Perturbative QCD}, World Scientific, Singapore (1989).
\bibitem{26} C. Adloff \textit{et al}. (H1 Collaboration), \textit{Eur. Phys. J.} \textbf{C21}, 33 (2001); \textbf{arXiv:}hep-ex/0012053.
\bibitem{27} J. Breitweg \textit{et al}. (ZEUS Collaboration), \textit{Phys. Lett.} \textbf{B487}, 53 (2000); \textbf{arXiv:}hep-ex/0005018.
\bibitem{28} P. Zavada, \textit{Phys. Rev.} \textbf{D83}, 014022 (2011).
\bibitem{29} M. Anselmino, M. Boglione, U. D'Alesio, A. Kotzinian, F. Murgia and A. Prokudin, \textit{Phys. Rev.} \textbf{D71}, 074006 (2005); \textbf{arXiv:}hep-ph/0501196.
\bibitem{30} A. Achilli, R. M. Godbole, A. Grau, R. Hegde, G. Pancheri and Y. Srivastava, \textit{Phys. Lett.} \textbf{B659}, 137 (2008); \textbf{arXiv:}0708.3626[hep-ph].
\bibitem{31} A. Achilli, Y. Srivastava, R. M. Godbole, A. Grau, G. Pancheri and O. Shekhovtsova, \textit{Phys. Rev.} \textbf{D84}, 094009 (2011); \textbf{arXiv:}1102.1949[hep-ph].
\bibitem{32} L. V. Gribov, E. M. Levin and M. G. Ryskin, \textit{Phys. Rep.} \textbf{100}, 1 (1983).
\bibitem{33} M. Gl{\"u}ck, E. Reya and A. Vogt, \textit{Euro. Phys. J.} \textbf{C5}, 461 (1998); \textbf{arXiv:}hep-ph
/9806404.
\bibitem{34} D. Ebert, R. N. Faustov and V. O. Galkin, \textit{Phys. Rev.} \textbf{D79}, 114029 (2009); \textbf{arXiv:}0903.5183[hep-ph].

\end{thebibliography}
\end{document}